\documentclass{ws-procs975x65}

\newcommand{\asize}{\mathcal{S}}
\newcommand{\SdS}{Schwarzschild--de~Sitter }

\begin{document}

\title{A VIRTUAL TRIP TO THE SCHWARZSCHILD--DE~SITTER BLACK HOLE\footnote{The present work was supported by the Czech grants MSM 4781305903 and LC06014 (P.B.)}}

\author{PAVEL BAKALA$^{\Delta}$, STANISLAV HLED\' IK$^{\nabla}$\\ ZDEN\v EK STUCHL\'IK$^{\star}$, KAMILA TRUPAROV\' A$^{\dag}$}

\address{Institute of Physics, Faculty of Philosophy and Science,Silesian University in Opava\\Bezru\v covo n\' am. 13, CZ-746\,01 Opava, Czech Republic\\
\email{$^{\Delta}$pavel.bakala@fpf.slu.cz, $^{\nabla}$stanislav.hledik@fpf.slu.cz\\$^{\star}$zdenek.stuchlik@fpf.slu.cz, $^{\dag}$kamila.truparova@fpf.slu.cz}}

\author{PETR \v CERM\' AK}

\address{Institute of Computer Science, Faculty of Philosophy and Science,
Silesian University in Opava\\Bezru\v covo n\' am. 13, CZ-746\,01 Opava, Czech Republic\\
\email{petr.cermak@fpf.slu.cz}}


\begin{abstract}
We developed realistic fully general relativistic computer code for simulation of optical projection in a strong, spherically symmetric gravitational field. Standard theoretical analysis of optical projection for an observer in the vicinity of a Schwarzschild black hole is extended to black hole spacetimes with a repulsive cosmological constant, i.e, \SdS (SdS) spacetimes. Influence of the cosmological constant is investigated for static observers and observers radially free-falling from static radius.
Simulation includes effects of gravitational lensing, multiple images, Doppler and gravitational frequency shift, as well as the amplification of intensity. The code generates images of static observers sky and a movie simulations for radially free-falling observers. Techniques of parallel programming are applied to get high performance and fast run of the simulation code.
\end{abstract}

\bodymatter

\section{Introduction}

Recent observations indicate the cosmological expansion accelerated by a dark energy that can be described by a repulsive cosmological constant, $\Lambda > 0$\cite{our_web}. We investigate influence of $\Lambda > 0$ on the appearance of distant universe for observers in close vicinity of the SdS black hole. Visualization outputs of the simulation code, images and movies, can be downloaded from our web site \cite{our_web}.

\section{Optical projection in \SdS spacetimes}
Construction of relativistic optical projection consists in finding all null geodesics connecting the source and the observer. In SdS spacetimes, the null geodesics are characterised by the impact parameter $b$, defined as the ratio of motion constants by the relation $b\equiv\Phi/\mathcal{E}$\cite{SH,SP}. The motion of photons in SdS spacetimes is governed by the Binet formula \cite{BCHST,SP}
\begin{equation}
\frac{d \phi}{du}=\pm \frac{1}{\sqrt{b^{-2}-u^{2}+2u^{3}+y}},
\label{Binet}
\end{equation}
where $u=r^{-1}$ and dimensionless cosmological parameter reads
$y =\textstyle\frac13\Lambda M^2$. Here $M$ is mass of the central black hole and $\Lambda \sim
{10}^{-56}\,\mathrm{cm}^{-2}$ is the repulsive cosmological constant \cite{SH,SP}.
The critical impact parameter $b_{\mathrm{c}} =\sqrt{27/\left(1-27y\right)}$ corresponds to the circular photon geodesic\cite{SH,SP}. Photons coming from distant universe with $b<b_{\mathrm{c}}$ finish in the central singularity, while photons with $b>b_{\mathrm{c}}$ return to the cosmological horizont\cite{BCHST,SP}.
An image will be observed in the direction determined by the space part of 4-momentum of photon, tangent to the photon trajectory. In order to obtain the optical projection for a given observer, the 4-momentum of the photons has to be transformed into observer's local frame. The solution expressing $b$ as a function of the source angular coordinate and of the image order\cite{BCHST} has to be done numerically.

\section{Apparent angular size of the black hole.}

\begin{figure}
\begin{minipage}[b]{.48\hsize}
\centering
\includegraphics[width=\hsize]{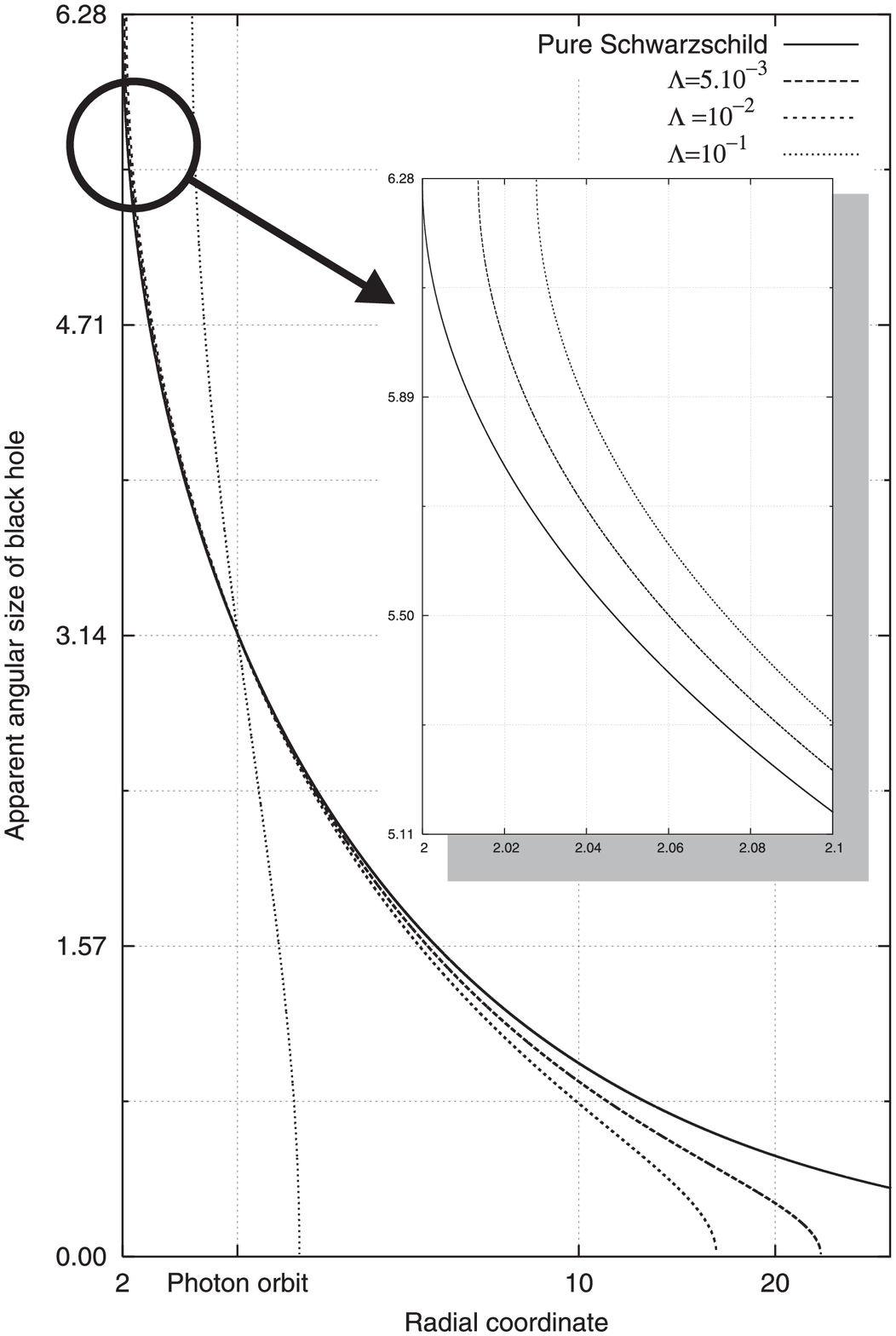}
\end{minipage}\hfill%
\begin{minipage}[b]{.48\hsize}
\centering
\includegraphics[width=\hsize]{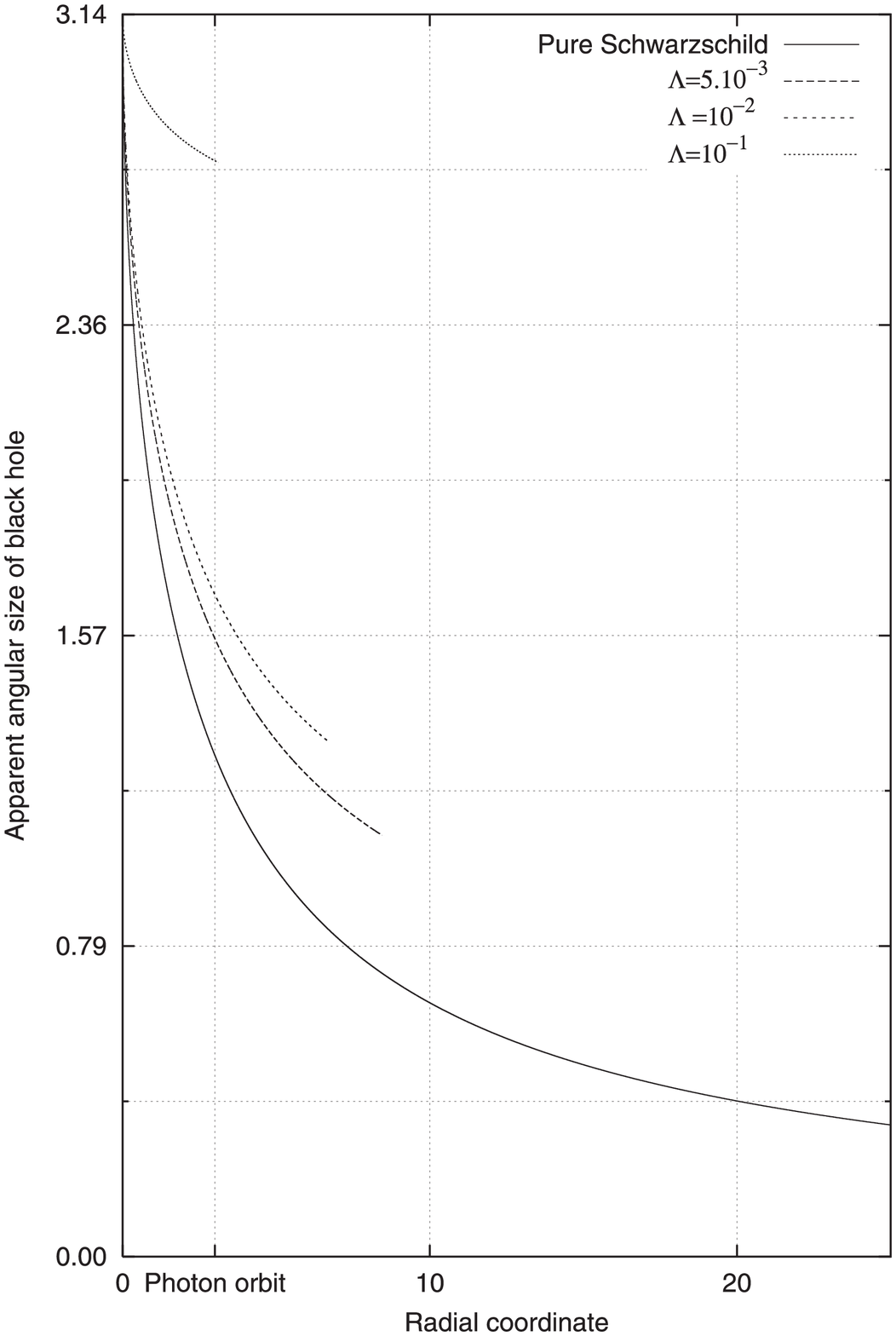}
\end{minipage}
\noindent
\caption{Apparent angular size of the black hole as function of observer's radial coordinate.
\textbf{Left panel:} Static observers. \textbf{Right panel:} Radially free-falling observers}

\label{F1}
\end{figure}


The apparent angular size $\asize$ of the black hole can be naturally defined as the observed angular size of the circular black region on the observer sky, in which no images of distant objects exist and only radiation originated under the circular photon orbit can be observed \cite{BCHST,SP,cunn01}.
For observers located above the circular photon orbit the boundary of the black region corresponds to the outgoing geodesics with  impact parameter approaching $b_{c}$ from above, while for observers located under the circular photon orbit the boundary corresponds to the ingoing geodesics with the impact parameter approaching $b_{c}$ from below.

In the case of static observers we have \cite{BCHST}
\begin{equation}
\asize = 2\arccos A(r_{obs},y;b) \qquad  \mbox{,}\qquad  A(r,y;b)\equiv \pm\sqrt{1-\frac{b^{2}}{r^{2}}\left(1-\frac{2}{r}-yr^2\right) }\,.
\label{asizestat}
\end{equation}
Here $+$' and `$-$'s signs correspond to observers located above and under the circular photon orbit, respectively.
Above the circular photon orbit increasing cosmological constant causes shrinking of the black region, whereas under the circular photon orbit the black region grows with increasing cosmological constant. In the limit case of observers located just on the circular photon orbit, $\asize$ is independent of the cosmological constant. It is invariably $\pi$, i.e., the black region always occupies just one half of the observer sky.

In  the case of observers radially free-falling from the static radius \cite{BCHST,SH,SP}, the  apparent angular size of the black hole reads \cite{BCHST}

\begin{equation}\asize = 2\arccos  \frac{\left (Z(r_{obs},y)+\sqrt{1-3y^{1/3}}A(r_{obs},y;b) \right )} {\left( \sqrt{1-3y^{1/3}}+Z(r_{obs},y)A(r_{obs},y;b) \right )},
\label{asizefall}
\end{equation}
where
\begin{equation}
Z(r,y) \equiv \sqrt{\frac{2}{r}+yr^2-3y^{1/3}}.
\end{equation}
Dependency on the cosmological constant is qualitatively different. For radially free-falling observers $\asize$ grows with increasing cosmological constant at all values of the radial coordinate except the central singularity, where $\asize$ is invariably $\pi$, similarly to the case of static observers located on the circular photon orbit. Consequently, the radially free-falling observer will always observe smaller $\asize$ then the static observer on the same radial coordinate.

\vfill


\begin{thebibliography}{20}

\bibitem{BCHST} P.Bakala, P. \v Cerm\' ak, S. Hled\' ik, Z. Stuchl\' ik, K. Truparov\' a  Pl\v skov\' a : A virtual trip to the Schwarzschild-de Sitter black hole. Proceedings of RAGtime 6/7: Workshop on blackholes and neutron stars, Opava, September 2004/2005

\bibitem{cunn01} C.T. Cunningham: Optical Appearance of Distant Observers near
  and inside a Schwarzschild Black Hole. Phys .Rev. D.12, 323--328, 1975.

\bibitem{nem01} R. J. Nemiroff: Visual distortion near a neutron star a and
  black hole. \url{arXiv:astro-ph/9312003}, 1993.

\bibitem{our_web} Relativistic and particle physics and its astrophysical aplications,
Czech research project MSM 4781305903
  \url{http://www.physics.cz/research/}.

\bibitem{SH} Z. Stuchl\' ik and. S. Hled\' ik : Some properties of the Schwarzschild--de~Sitter and Schwarzschild--anti--de~Sitter spacetimes.  Phys. Rev. D, 60(4):0044006(15 pages), 1999

\bibitem{SP} Z. Stuchl\' ik and K. Pl\v skov\' a : Optical apperance of isotropically radiating sphere in the Schwarzschild--de~Sitter spacetime.  Proceedings of RAGtime4/5 , 167--185, 2004

\end{thebibliography}
\end{document}